\documentclass[aps,prl,twocolumn,superscriptaddress,longbibliography]{revtex4-2}
\usepackage{amsmath,amssymb}
\usepackage[pdftex]{hyperref,graphicx}
\hypersetup{colorlinks = true, urlcolor = blue, linkcolor = blue, citecolor = blue}
\usepackage{physics}
\usepackage{xcolor}
\usepackage{bm}
\usepackage{pdfpages}

\newcommand{\comment}[1]{{}}

\makeatletter
\AtBeginDocument{\let\LS@rot\@undefined}
\makeatother

\begin{document}
\title{Hydrodynamics, viscous electron fluid, and Wiedeman-Franz law in 2D semiconductors}
\author{Seongjin Ahn}
\affiliation{Condensed Matter Theory Center and Joint Quantum Institute, Department of Physics, University of Maryland, College Park, Maryland 20742, USA}
\author{Sankar Das Sarma}
\affiliation{Condensed Matter Theory Center and Joint Quantum Institute, Department of Physics, University of Maryland, College Park, Maryland 20742, USA}

\begin{abstract}
Considering theoretically the transition between hydrodynamic and ballistic regimes in 2D semiconductors, we show that electrons in high-mobility 2D GaAs are by far the best system for the direct observation of collective hydrodynamic effects even in bulk transport properties independent of complicated transport features in narrow constrictions and small systems where Gurzhi phenomena are typically studied experimentally.  We predict a strong hydrodynamics-induced generic violation of the Wiedeman-Franz law in bulk 2D GaAs systems for mobilities as modest as $10^6 \mathrm{cm}^2/Vs$ and densities $1$-$5\times10^{11} \mathrm{cm}^{-2}$ in the temperature range of $T=1$-$40K$.

\end{abstract}

\maketitle
{\em Introduction.}--- 
It has been known for a long time \cite{pines2018theory}
%[cite the book by Pines-Nozieres “Theory of Quantum Liquids”] 
that an interacting Fermi liquid undergoes a crossover from ballistic (or collisionless or diffusive) regime to a collision-dominated collective hydrodynamic (or viscous) regime as the inelastic momentum-conserving electron-electron interaction strength, $1/\tau_{ee}$, increases, surpassing the momentum-relaxing elastic scattering rate, $1/\tau_\mathrm{e}$, arising from impurity and phonon scattering. In the limit, $\tau_\mathrm{ee}\gg\tau_\mathrm{e}$, the electrons flow ballistically as individual quasiparticles as there is no local thermodynamic equilibrium, but for $\tau_\mathrm{ee}\ll\tau_\mathrm{e}$, the electrons behave collectively as a fluid following continuum macroscopic laws of hydrodynamics since fast electron-electron scattering produces local thermodynamic equilibrium \cite{lucasHydrodynamicsElectronsGraphene2018}.  Defining a dimensionless `effective Knudsen' parameter $\zeta=\tau_\mathrm{ee}/\tau_\mathrm{e} = l_\mathrm{ee}/l_\mathrm{e}$ where $l_\mathrm{ee}$ ($l_\mathrm{e}$) is the inelastic momentum-conserving  (elastic momentum-relaxing) mean free path, hydrodynamics (ballistic) flow happens for $\zeta \ll (\gg) 1$ with a crossover around $\zeta\sim 1$.  The corresponding flow parameter in classical fluids is the Knudsen number $K_n$ defined as the ratio of the molecular collisional mean free path and the linear size of a physical object (e.g., constriction or obstruction). Large $K_n$, which can only happen in dilute gases, implies a failure of continuum fluid dynamics (implying a ballistic flow of the molecules).  In quantum electron fluids, the situation is richer since in principle several independent variables, $l_\mathrm{ee}$ and $l_\mathrm{e}$ as well as a physical constriction or obstruction size ($d$) can be varied to control the nature of the flow, creating many interesting experimental possibilities.  The mean free paths can be controlled in electronic systems by varying carrier density ($n$), temperature ($T$), and the amount of disorder (low-$T$ mobility, $\mu$).

It was pointed out by Gurzhi a long time ago \cite{gurzhiMinimumResistanceImpurityfree1963, gurzhiHydrodynamicEffectsSolids1968}
%[ Gurzhi, R. N. (1963). "Minimum of resistance in impurity-free conductors". J Exp Theor Phys. 17: 521. Gurzhi, R. N. (1968);   "HYDRODYNAMIC EFFECTS IN SOLIDS AT LOW TEMPERATURE". Soviet Physics Uspekhi. 11 (2): 255–270. doi:10.1070/PU1968v011n02ABEH003815]
that if the condition $l_\mathrm{e} \gg d \gg l_\mathrm{ee}$ is satisfied in a metal in a constrained geometry, then many unexpected and counter-intuitive hydrodynamics-induced transport phenomena (e.g., resistance decreasing with increase temperature) may arise.  But the condition $l_\mathrm{e} \gg l_\mathrm{ee}$ is essentially impossible to achieve in regular 3D metals at any temperatures, and the subject remained dormant for almost 50 years in spite of a report of the claimed observation of the Gurzhi effect in constrained GaAs systems 30 years ago \cite{dejongHydrodynamicElectronFlow1995}. 
%[de Jong, M. J. M.; Molenkamp, L. W. (1995). "Hydrodynamic electron flow in high-mobility wires". Phys. Rev. B. 51 (19): 13389–13402. arXiv:cond-mat/9411067. doi:10.1103/PhysRevB.51.13389]  
During the last 6-7 years, however, there have been several reports of the experimental observation of various aspects of the hydrodynamic effects in electron liquids, mostly in clean graphene layers \cite{krishnakumarSuperballisticFlowViscous2017, bandurinFluidityOnsetGraphene2018}
%[https://www.nature.com/articles/nphys4240; https://doi.org/10.1038/s41467-018-07004-4]
but also in other materials \cite{goothThermalElectricalSignatures2018, mollEvidenceHydrodynamicElectron2016, scaffidiHydrodynamicElectronFlow2017}.
%[https://www.science.org/doi/10.1126/science.aac8385; https://journals.aps.org/prl/abstract/10.1103/PhysRevLett.118.226601]  
In the current work, we focus on 2D GaAs systems, establishing that it is almost an ideal system for studying electron hydrodynamics by calculating its effective Knudsen number over a large range of density and temperature.  There has been some experimental work on hydrodynamic aspects of 2D GaAs, mostly on various Gurzhi effects in geometrically constrained systems (with a finite $d$) with constrictions and obstructions \cite{gusevViscousMagnetotransportGurzhi2021, pusepDiffusionPhotoexcitedHoles2022, gusevViscousElectronFlow2018, keserGeometricControlUniversal2021, guptaHydrodynamicBallisticTransport2021}.
%[ https://doi.org/10.1103/PhysRevB.103.075303; https://doi.org/10.1103/PhysRevLett.128.136801; https://aip.scitation.org/doi/10.1063/1.5020763; https://doi.org/10.48550/arXiv.2103.09463; https://doi.org/10.1103/PhysRevLett.126.076803]
But our focus is a bulk 2D electron gas, with no geometric constraint or constriction, i.e., $d$ is effectively infinite.  Through concrete theoretical calculations we show explicitly that the effective Knudsen parameter in readily available 2D GaAs is much smaller than unity rather generically, $\zeta\ll1$, over a large range of temperature and carrier density, thus making the GaAs-based 2D electron system a generic viscous hydrodynamic fluid, without any constriction or obstruction or boundary effects, leading to a strong violation of the Wiedeman-Franz (WF) law associated with its highly viscous nature.  The 2D GaAs structures are thus generically WF law violating Fermi liquids, in spite of the common knowledge that GaAs is a weakly interacting system where the electron interaction parameter $r_s$, the ratio of the Coulomb interaction energy to the noninteracting kinetic energy, is typically small ($r_s <1$), compared with 3D metals where $r_s \sim 6$.  We mention that other 2D semiconductor systems, e.g., Si 100 inversion layers, although much more strongly interacting ($r_s \sim 3$-$12$), are never in the hydrodynamic regime with $\zeta>1$.

\begin{figure}[!htb] 
    \centering
    \includegraphics[width=1.0\linewidth]{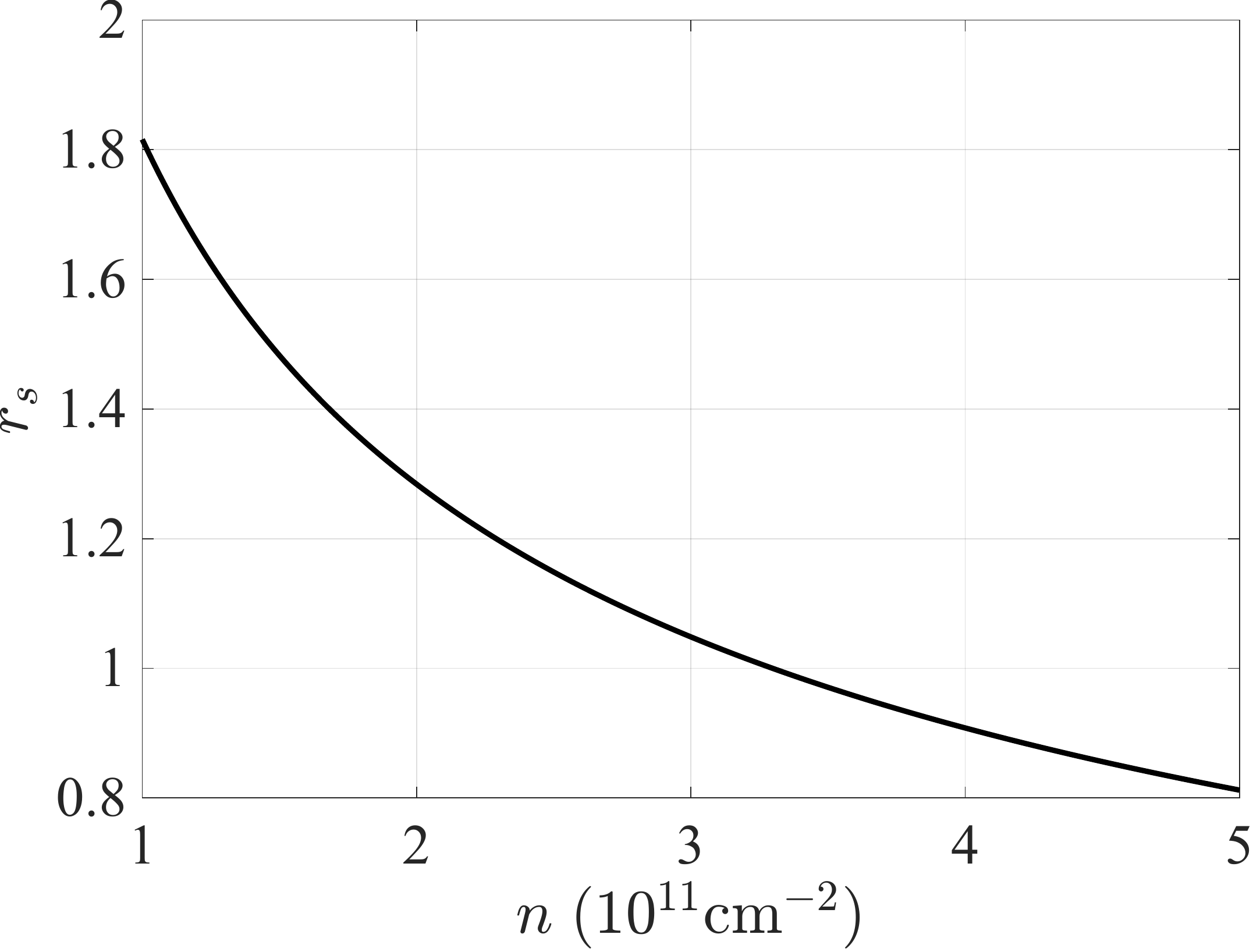}
    \caption{Shows the calculated interaction parameter as a function of carrier density in 2D GaAs}
    \label{fig:1}
\end{figure}

{\em Theory and Results.}--- 
The momentum relaxation rate in a 2D semiconductor is characterized by its measured electron mobility, and in 2D GaAs system the typical mobility is $\sim 10^6$-$10^7 \mathrm{cm}^2/Vs$, with the reported state-of-the-arts highest achieved mobility being $> 5\times10^7 \mathrm{cm}^2/Vs$ \cite{chungUnderstandingLimitsMobility2022, chungUltrahighqualityTwodimensionalElectron2021a, ahnDensitydependentTwodimensionalOptimal2022a, hwangLimitTwodimensionalMobility2008a}. 
%[https://arxiv.org/abs/2206.13902; https://www.nature.com/articles/s41563-021-00942-3; https://journals.aps.org/prmaterials/abstract/10.1103/PhysRevMaterials.6.014603; https://journals.aps.org/prb/abstract/10.1103/PhysRevB.77.235437]   
The measured mobility, $\mu$, is directly related to the momentum relaxation rate and the elastic mean free path $l_e$ using the known GaAs electron effective mass ($m=0.07 m_e$) and the 2D carrier density $n$:
\begin{equation}
    l_e = 5.22 \mu n^{1/2}
    \label{eq:elastic_mean_free_path}
\end{equation}
where $n$ is measured in units $10^{11} \mathrm{cm}^{-2}$, $\mu$ in units of $10^6 \mathrm{cm}^2/Vs$, and $l_e$ in $\mu m$ in Eq.~(\ref{eq:elastic_mean_free_path}). Thus, ten million mobility, which is routine (and was achieved already in 1990) \cite{stormerObservationBlochGruneisenRegime1990},
%[https://journals.aps.org/prb/abstract/10.1103/PhysRevB.41.1278] 
corresponds to an elastic mean free path of $\sim 52$ microns for $n\sim 10^{11}$ density, and proportionally higher at higher densities. Such exceptionally long macroscopic elastic mean free path may misleadingly imply that the electrons travel basically ballistically through a 2D GaAs sample, suffering impurity collisions rarely. But as we show below, the reverse is in fact true: the electron flow through a bulk high-mobility 2D GaAs sample is in fact `sluggish', being a highly viscous bulk hydrodynamic flow (much more viscous than the flow of honey!)  due to the dominant role of electron-electron scattering.

The momentum conserving scattering rate due to electron-electron interaction can be perturbatively calculated by obtaining the imaginary part of the finite-temperature electron self-energy using diagrammatic many body theory, and has recently been calculated up to the next-leading-order in temperature \cite{liaoTwodimensionalElectronSelfenergy2020a, dassarmaKnowEnemy2D2021a}
%[https://journals.aps.org/prb/abstract/10.1103/PhysRevB.102.085145; https://doi.org/10.1016/j.aop.2021.168495] 
and leading order in $r_s$.  The analytical calculations are quite intricate and here we simply quote the results:
\begin{equation}
    \begin{split}
        \mathrm{Im}\Sigma^\mathrm{(R)}(|\varepsilon|\ll T) =& -\frac{\pi}{8} \frac{T^2}{E_\mathrm{F}} \ln\left(\frac{\sqrt{2}r_s E_\mathrm{F}}{T} \right) \\
        &+ \frac{\pi}{24}(6 + \ln(2\pi^3) - 36 \ln A)\frac{T^2}{E_\mathrm{F}} \\
        &- \frac{7\zeta(3)}{2\sqrt{2}\pi}\frac{T^3}{r_s E_\mathrm{F}^2}
    \end{split}
    \label{eq:self-energy}
\end{equation}
The right hand side of Eq.~(\ref{eq:self-energy}) gives us $1/2 \tau_\mathrm{ee}$ as a function of $T$ and $n$ through $r_s = (\pi a_B^2 n)^{-1/2}$ and $E_\mathrm{F}=2\pi n/gm$, where $a_B=\kappa/me^2$ is the effective Bohr radius, $m$ is the electron mass, $\kappa$ is the dielectric constant, and $g=2$ is the spin degeneracy.  Here we set $\hbar=1$.
Note that the correct result up to the leading order $T^2$ term was first obtained in \cite{zhengCoulombScatteringLifetime1996a, murphyLifetimeTwodimensionalElectrons1995a}. 
%[https://journals.aps.org/prb/abstract/10.1103/PhysRevB.53.9964]
Writing $E_\mathrm{F} =T_\mathrm{F}$ with $k_\mathrm{B} = 1$, we immediately see that the inelastic scattering rate is strongly suppressed for $T\ll T_\mathrm{F}$ i.e., at low $T$ and/or large densities.  Unfortunately, Eq.~(\ref{eq:self-energy}) is of little use to us since it is valid at asymptotically low temperatures where the electron-electron scattering rate is low, and hydrodynamics does not apply by definition.  We therefore directly numerically calculate the momentum-conserving mean free path $l_\mathrm{ee}$ by calculating the imaginary part of the $T$-dependent electron self-energy, summing the infinite series of bubble diagrams (i.e. the RPA series), which is exact in the high-density limit, and is known to produce reasonable quantitative results as long as  $r_s$  is not too large, which it is not for 2D GaAs (see Fig.~\ref{fig:1}) in the density range of our interest.

\begin{figure*}[!htb] 
    \centering
    \includegraphics[width=1.0\linewidth]{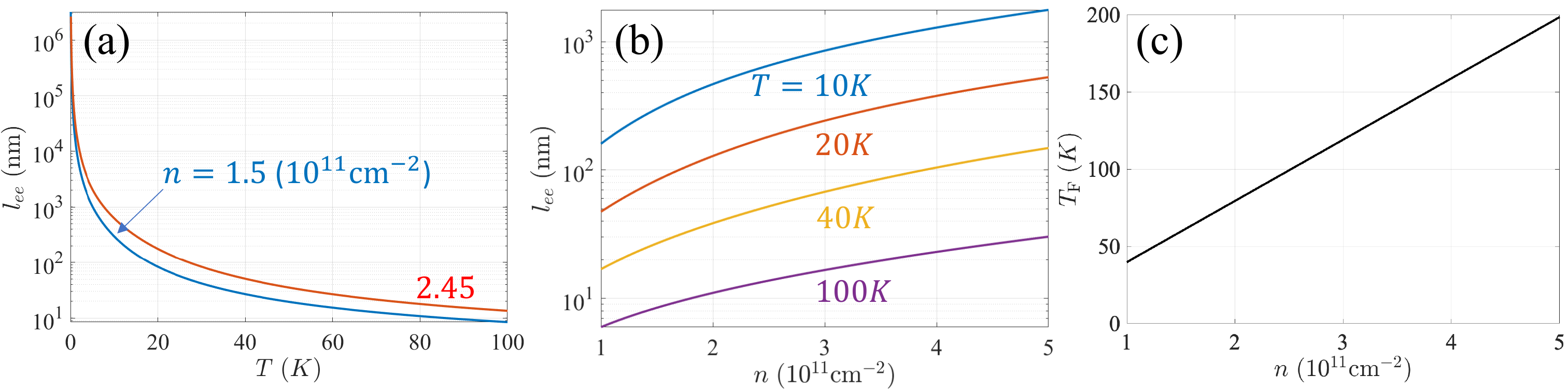}
    \caption{Calculated inelastic mean free path in 2D GaAs as a function of (a) $T$ for $n=1.45$ and $2.5\times10^{11} \mathrm{cm}^{-2}$ density and as a function of (b) $n$ for $T=10$, $20$, $40$ and $100$K.  Panel (c) shows the calculated Fermi temperature $T_\mathrm{F}$ as a function of $n$ for 2D GaAs}
    \label{fig:2}
\end{figure*}

The inelastic mean free path is given by $l_\mathrm{ee} = v_\mathrm{F} \tau_\mathrm{ee}$, where
\begin{equation}
    \hbar/\tau_\mathrm{ee}(T)=2\mathrm{Im}\Sigma(k_\mathrm{F},\xi_{k_\mathrm{F}},T)
\end{equation}
and, in the RPA theory of the infinite sum of bubble diagrams:
\begin{align}
    \mathrm{Im}\Sigma(\bm k, \omega,T)
    \!=&\!\int\!\frac{d^2 q}{(2\pi)^2} \left [n_\mathrm{B}(\hbar\omega-\xi_\mathrm{\bm k+\bm q}) + n_\mathrm{F}(-\xi_\mathrm{\bm k+\bm q}) \right ] \nonumber \\ 
&\times v_c(\bm q)\mathrm{Im}\left[\frac{1}{\varepsilon(\bm q,\xi_\mathrm{\bm k+\bm q}-\hbar\omega, T)}\right],
    \label{eq:imag_self_energy}
\end{align}
where $n_\mathrm{F(B)}$ is the Fermi(Bose) distribution function and $\xi=k^2/2m-\mu(T)$ with $\mu(T)$ denoting the chemical potential. Note that the Coulomb interaction $v_c(q)$ is dynamically screened, divided by the dynamic dielectric function given by $\varepsilon(q, \omega,T)=1-v_c(q)\Pi(q,\omega,T)$, where $\Pi$ is the finite-T 2D polarizability (i.e., the bare bubble) function.  We carry out the 3D integration for Eq.~(\ref{eq:imag_self_energy}) using the appropriate 2D GaAs parameters, where one integration is necessary to obtain the finite-$T$ polarizability, as well as the self-consistent calculation of the chemical potential $\mu(T)$, numerically in order to calculate $l_\mathrm{ee}$ which is shown in Fig.~\ref{fig:2}.  Our numerical results agree with the analytical results of Eq.~(\ref{eq:imag_self_energy}) for low $T < 0.1 T_\mathrm{F}$.

\begin{figure*}[!htb] 
    \centering
    \includegraphics[width=1.0\linewidth]{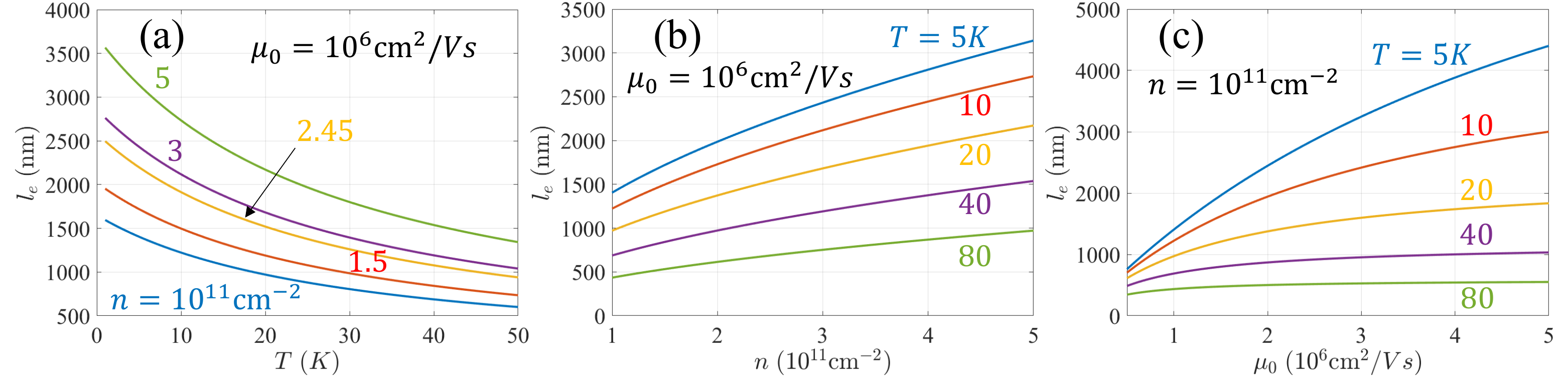}
    \caption{Calculated elastic mean free path as a function of (a) $T$ including acoustic phonon scattering for 5 densities and as a function of (b) $n$ for 5 temperatures. The zero-$T$ impurity scattering induced mobility is $10^6  \mathrm{cm}^2/Vs$ for panels (a) and (b).  In panel (c), the calculated $l_e$ is shown as a function of $T=0$ mobility at a fixed $n=10^{11} \mathrm{cm}^{-2}$ for 5 temperatures (only acoustic phonon scattering included).}
    \label{fig:3}
\end{figure*}

A direct comparison between Fig.~\ref{fig:2} and Eq.~(\ref{eq:elastic_mean_free_path}) shows that the effective electronic Knudsen parameter $\zeta = l_\mathrm{ee}/l_\mathrm{e}<1$ already for $T>10$K and $n<5\times10^{11} \mathrm{cm}^{-2}$ and a modest $\mu= 10^6 \mathrm{cm}^2/Vs$.  This implies a large regime of viscous bulk hydrodynamic electronic flow in 2D GaAs, generically accessible to all experiments without using any constrictions, boundaries, and obstructions. There is, however, one catch—we must include electron-phonon scattering in the theory to recalculate the finite-$T$ mobility and the associated $l_\mathrm{e}(T)$ since phonon scattering would suppress both the mobility and the momentum relaxation mean free path from its low-$T$ values quoted in experiments. Particularly, for $T>40$K, LO-phonon scattering becomes important in GaAs, leading to an exponential (in $T$) decrease in the mobility (and consequently $l_\mathrm{e}$) \cite{kawamuraPhononscatteringlimitedElectronMobilities1992},
%[https://journals.aps.org/prb/abstract/10.1103/PhysRevB.45.3612] 
 leading to $\zeta=l_\mathrm{ee}/l_\mathrm{e}>1$, recovering ballistic flow for $T>40$K.  Below the LO-phonon scattering regime, acoustic phonon scattering suppresses mobility in the equipartition regime of quasi-elastic scattering for $T> T_\mathrm{BG}/6$, where $T_\mathrm{BG} = 2\hbar s k_\mathrm{F} \sim n^{1/2}$ is the Bloch-Gruneisen temperature ($s$ is the speed of sound and $k_\mathrm{F}$ the Fermi wavenumber) \cite{kawamuraPhononscatteringlimitedElectronMobilities1992,hwangLimitTwodimensionalMobility2008a, minInterplayPhononImpurity2012a}
%[https://journals.aps.org/prb/abstract/10.1103/PhysRevB.45.3612; https://journals.aps.org/prb/abstract/10.1103/PhysRevB.77.235437; https://journals.aps.org/prb/abstract/10.1103/PhysRevB.86.085307]
(Below this characteristic temperature, acoustic phonon scattering is strongly suppressed and is irrelevant for our considerations.) We calculate the acoustic phonon scattering induced finite-$T$ correction to the momentum relaxation mean free path $l_e$ including both piezoelectric coupling and deformation potential coupling in the equipartition quasielastic regime by using parameters appropriate for 2D GaAs and following Refs.~\cite{kawamuraPhononscatteringlimitedElectronMobilities1992, minInterplayPhononImpurity2012a}.
%[https://journals.aps.org/prb/abstract/10.1103/PhysRevB.45.3612; https://journals.aps.org/prb/abstract/10.1103/PhysRevB.86.085307]  
The mobility and the mean free path decrease as $1/T$ due to the phonon corrections in this ``high-$T$'' (i.e., $T>T_\mathrm{BG}/6$) regime.  The calculated result including phonon scattering is shown in Fig.~\ref{fig:3}.

\begin{figure}[!htb] 
    \centering
    \includegraphics[width=1.0\linewidth]{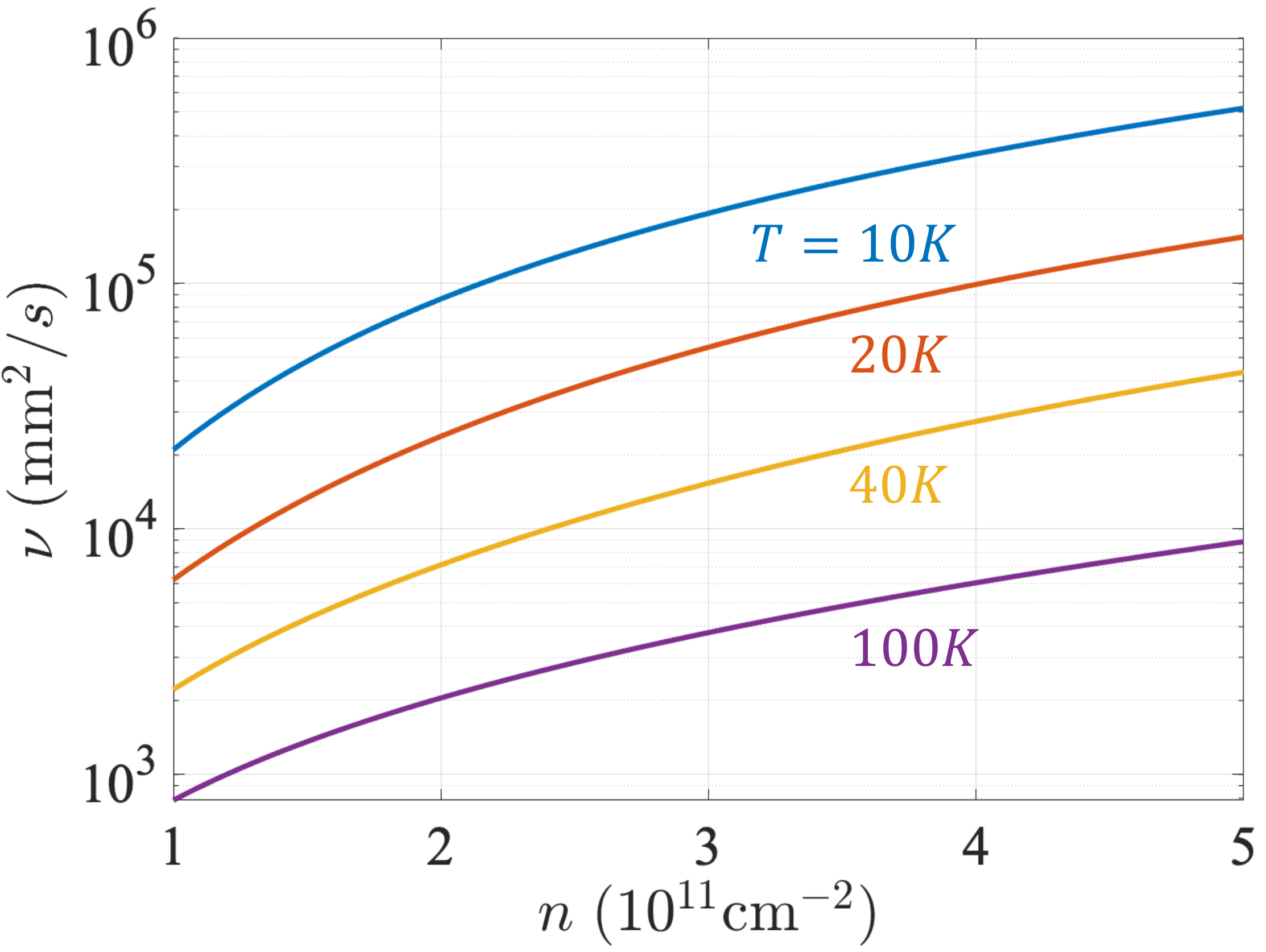}
    \caption{Calculated viscosity in 2D GaAs for $T=10$, $20$, $40$, and $100$K.}
    \label{fig:viscosity}
\end{figure}

A comparison of Figs.~\ref{fig:2} and \ref{fig:3} manifestly show that generically the Knudsen parameter in 2D GaAs is $\zeta \ll 1$ in large regimes of 2D density ($n\sim 1$-$5\times10^{11}\mathrm{cm}^{-2}$) and temperature ($T \sim 5$-$40$K), making viscous bulk hydrodynamic flow the generic property of 2D GaAs electrons for mobilities as low as $10^6 \mathrm{cm}^2/Vs$.
A rough estimate of the bulk kinematic viscosity coefficient of GaAs electrons is given by $\nu\sim v_\mathrm{F} l_\mathrm{ee}$, provided that the hydrodynamic limit of $l_\mathrm{e} \gg l_\mathrm{ee}$ applies, where $v_\mathrm{F}$, the Fermi velocity, depends on the density $n$ through $v_\mathrm{F}= 1.4n^{1/2} 10^8 \mathrm{mm/s}$ where $n$ is expressed in units of $10^{11} \mathrm{cm}^{-2}$.  Using our calculated $l_\mathrm{ee}$, we estimate $\nu$ to be $\sim 2\times10^4 \mathrm{mm}^2/s$ at $T=20$K and $n=1.5\times10^{11} \mathrm{cm}^{-2}$ (increasing almost to $10^6$ at higher densities) for 2D GaAs (see Fig.~\ref{fig:viscosity}), to be compared with the corresponding kinematic viscosity of honey (mayonnaise) $\sim 2(6)\times10^3 \mathrm{mm}^2/s$ at room temperatures.
We know of no other system, 2D or 3D (and metals or semiconductors or semimetals, including graphene) where the electron hydrodynamics manifests as clearly and as generically as in 2D GaAs.

\begin{figure}[!htb] 
    \centering
    \includegraphics[width=1.0\linewidth]{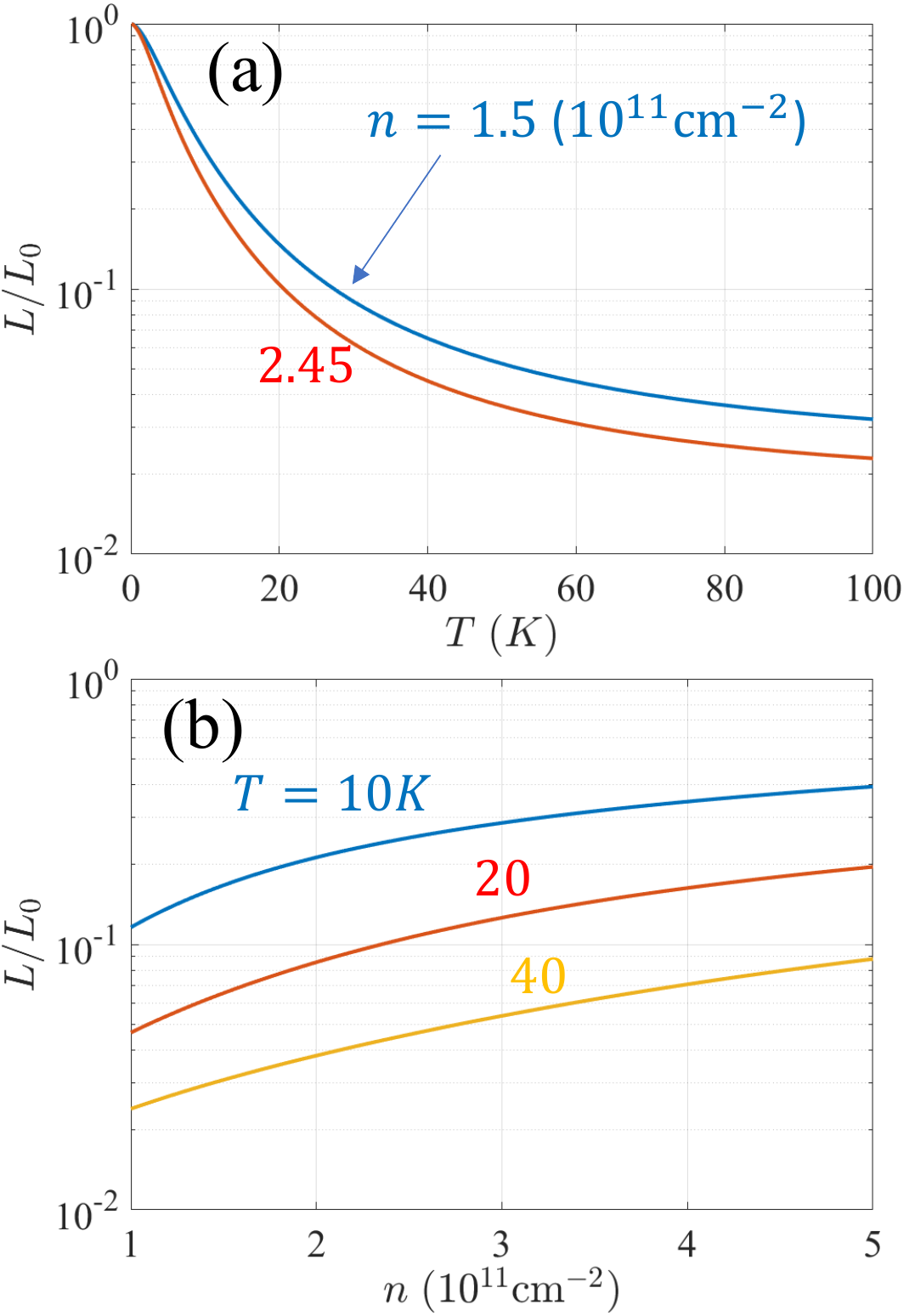}
    \caption{ Shows the calculated WF ratio (for a $T=0$ mobility of $10^6 \mathrm{cm}^2/Vs$) as a function of  (a) temperature for 2 carrier densities and as a function of (b) density for 3 temperatures.}
    \label{fig:5}
\end{figure}
Finally, we provide one spectacular example of hydrodynamic effects on a particular property of the GaAs 2D electron system in Fig.~\ref{fig:5}, where we calculate the WF ratio, $L/L_0$ for 2D GaAs, where
$L_0=(\pi^2/3)\times(k_\mathrm{B}/e)^2$ is the ideal free-electron Lorenz number for $\kappa/(\sigma T)$, with $\kappa$ and $\sigma$ are the electronic thermal and electrical conductivity, respectively.  It is well-known that all metals and Fermi liquids approximately obey $L/L_0 \sim 1$ in spite of Fermi liquid interactions, and often the validity (invalidity) of the WF law is identified with the existence of a Fermi (non-Fermi liquid).  We follow Refs.~\cite{lavasaniWiedemannFranzLawFermi2019, lucasElectronicHydrodynamicsBreakdown2018} %[https://journals.aps.org/prb/abstract/10.1103/PhysRevB.99.085104; https://journals.aps.org/prb/abstract/10.1103/PhysRevB.97.245128] 
in calculating $L/L_0$ from our calculated $l_\mathrm{e}$ and $l_\mathrm{ee}$ through the simple but exact formula:  
\begin{equation}
    L/L_0 = \frac{l_\mathrm{ee}}{l_\mathrm{ee} + l_\mathrm{e}}= \frac{\zeta}{1+\zeta}
\end{equation}
The spectacular failure of the WF law, with an order of magnitude suppression of the WF ratio is obvious already at $T\sim 10$K for $n\sim 10^{11}\mathrm{cm}^{-2}$.  Such a dramatic failure of the WF law, as predicted in our work, would have immediately led to the discussion of a possible breakdown of the Fermi liquid theory (even if $r_s$ is small), but it is in fact a spectacular example of the effect of bulk hydrodynamics, entirely within the Fermi liquid paradigm, arising simply from the fact that the 2D GaAs electrons become super-viscous as the momentum-conserving mean free path arising from electron-electron collisions become much shorter than the momentum-relaxing mean free path arising from disorder and phonon scattering.

{\em Conclusion.}--- 
We establish 2D GaAs electron systems as the ideal laboratory system to study viscous hydrodynamic effects in bulk electron fluids without invoking any finite size constrictions or obstructions or imposed geometric controls.  All one needs is to measure the Wiedeman-Franz ratio in 2D GaAs electrons in the temperature range of $T=4$-$40$K using samples of modest mobilities $\sim 10^6 \mathrm{cm}^2/Vs$, and we predict a spectacular suppression of the WF ratio.  We emphasize, however, that the system is still a Fermi liquid, and therefore the WF ratio will eventually go to unity at low (high) enough temperatures (below 1K and above 40K).  In fact, the recovery of the WF ratio to unity at low enough temperatures is a generic property of all non-Fermi-liquids \cite{Berg}, 
%[Erez Berg, private communications and to be published]
and any attempt to distinguish Fermi liquids and non-Fermi liquids based on the measured WF ratio is doomed to failure since it is incapable of distinguishing between hydrodynamics and non-Fermi liquids.
\section{Acknowledgement} \label{sec:acknowledgement}
This work is supported by the Laboratory for Physical Sciences.


% \bibliography{ref}

\end{document}